# Mechanistic Origin of Charge Separation and Enhanced Photocatalytic Activity in D–π–A-Functionalized UiO-66-NH$_2$ MOFs

Anastasiia Kultaeva[1*], Volodymyr Vasylkovskyi[1], Andreas Sperlich[1], Eugenio Otal[2], Katsuya Teshima[2,3,4], Wolf Gero Schmidt[5], Timur Biktagirov[5*]

[1]*Experimental Physics 6 and Würzburg-Dresden Cluster of Excellence ct.qmat, Julius-Maximilian University of Würzburg, 97074 Würzburg, German*
[2]*Institute for Aqua Regeneration (ARG), Shinshu University, 4-17-1 Wakasato, Nagano 380-8553, Japan*
[3]*Department of Materials Chemistry, Faculty of Engineering, Shinshu University, 4-17-1 Wakasato, Nagano 380-8553, Japan*
[4]*Research Initiative for Supra-Materials (RISM), Interdisciplinary Cluster for Cutting Edge Research (ICCER), Shinshu University, Nagano 380-8553, Japan*
[5]*Physics Department, Paderborn University, D-33098 Paderborn, Germany*

*Corresponding authors

## Abstract

Donor–π–acceptor (D–π–A) functionalization of MOF linkers can enhance visible-light photocatalytic activity, yet the mechanisms responsible for these effects remain unclear. Here we combine EPR spectroscopy, transient photoluminescence, and first-principles calculations to examine how diazo-coupled anisole, diphenylamine (DPA), and N,N-dimethylaniline (NNDMA) groups modify the photophysics of UiO-66-NH$_2$. All donor units introduce new occupied states near the valence-band edge, enabling charge separation through dye-to-framework electron transfer. Among them, the anisole-modified material stands out for facilitating efficient intersystem crossing into a triplet charge-transfer configuration that suppresses fast recombination and yields long-lived charge carriers detectable by photo-EPR. Meanwhile, bulkier donors such as DPA and NNDMA – despite their stronger electron-donating character – also tend to introduce defect-associated trap states. These results underscore the interplay between donor-induced electronic-structure changes, triplet pathways, and defect-mediated recombination, offering a mechanistic basis for tuning photocatalytic response in D–π–A-modified MOFs.

## Introduction

Metal-organic frameworks (MOFs) have emerged as a versatile class of crystalline porous materials with tunable structures and functionalities, finding applications in areas such as catalysis, gas storage, and sensing [1, 2]. A notable example is the Zr-based UiO-66 framework [3, 4], which features Zr$_6$O$_4$(OH)$_4$ clusters linked by terephthalate units and is renowned for exceptional chemical and thermal stability [5] allowing the inclusion of different linker modifications [6]. UiO-66 and its derivatives have shown potential in photocatalysis. However, the parent material is a wide-band-gap semiconductor, absorbing mainly in the ultraviolet (UV) range (with a peak around 300 nm) [7, 8] – a significant drawback for solar-driven applications.

One strategy to extend the light absorption of UiO-66 into the visible range is to introduce electron-donating functional groups on its linkers, which introduce new filled (or occupied) states within the band gap and red-shift the absorption edge [6, 8–10]. For example, amino-functionalization reduces the band gap from ~4.0 eV in UiO-66 to ~2.9 eV in UiO-66-NH$_2$ [9].

An effective strategy to achieve further enhancement is through the incorporation of organic dyes via post-synthetic modifications [11]. Recent work has shown that post-synthetic covalent modification of UiO-66-NH$_2$ via diazonium coupling can shift the absorption into the visible spectrum [12]. This approach exploits the amino groups of UiO-66-NH$_2$ as anchoring points, by transforming them into diazo compounds (–N≡N$^+$) that react with the aromatic moiety of dye molecules forming stable azo (–N=N–) linkages within the MOF structure [13] (as illustrated in Figure 1A,B). While grafting chromophore units onto MOF linkers is recognized as a promising route toward more efficient light harvesting, a detailed understanding of the photophysical consequences is needed to rationalize and optimize the improvements.

In this study, we focus on covalent modification of the UiO-66-NH$_2$ linkers that follows the design principles of donor–π–acceptor (D–π–A) dyes [14]. Azo dyes with D–π–A architecture contain a –N=N– azo bridge connecting an electron-donor aryl group to an electron-acceptor moiety. Such dyes are widely used as photosensitizers to extend the light absorption of photocatalysts (for instance, grafted on TiO$_2$) [15] and can be tuned by varying their donor, bridge, or acceptor components [16]. A key principle in D–π–A design is to localize the highest occupied molecular orbital (HOMO) on the donor segment and the lowest unoccupied molecular orbital (LUMO) on the acceptor [17, 18]. This promotes intramolecular charge transfer upon photoexcitation, facilitating charge separation: the photoexcited electron shifts from the donor to the acceptor segment and can be then injected into the conduction band of a semiconductor photocatalyst. Implementing this concept in a MOF, we expect that an appended donor group can act as a visible-light absorber and funnel electrons into the MOF's inorganic nodes, improving light harvesting and photocatalytic activity. By combining experimental and theoretical results, we examine how the modifications influence light absorption, charge separation, and intersystem crossing in the dye-functionalized MOFs and derive insights for the rational design of MOF-based photocatalysts.

# Results and Discussion

**Post-synthetic Modifications, Optical Properties and Photocatalytic Activity**

We investigate UiO-66-NH$_2$ (Figure 1A) functionalized with three representative donor aryl groups of increasing electron-donating strength (see Figure 1B): anisole (–OC$_6$H$_5$CH$_3$), N,N-dimethylaniline (–C$_6$H$_4$N(CH$_3$)$_2$), and diphenylamine (–C$_6$H$_4$NHC$_6$H$_5$). These molecules were covalently attached to the 2-aminoterephthalate linkers of UiO-66-NH$_2$ via diazo coupling (forming –N=N– linkages). For brevity, we refer to the resulting MOFs as UiO-66-NH$_2$–anisole, –NNDMA, and –DPA, respectively. All three organic groups contain strong electron-donating substituents (methoxy, dimethylamino, and diphenylamino), as reflected by their Hammett σ constants in the para position [19]. The terephthalate linker (tethered to Zr nodes) plays the electron acceptor role due to its electron-withdrawing carboxylate groups.

Visually, the MOF powders changed color upon functionalization (from pale yellow for UiO-66-NH$_2$ to orange or reddish-brown for the dye-functionalized samples), qualitatively indicating successful dye incorporation and enhanced visible light absorption (Supplementary

Figure S1). The UV–Vis spectra shown in Figure 1C confirm that all three functionalized MOFs have significantly expanded absorption into the visible region compared to the parent UiO-66-NH$_2$. The unmodified UiO-66-NH$_2$ shows a steep absorption edge ~430 nm (consistent with its ~2.9 eV band gap [9]). Upon grafting the donor groups, the absorption onsets are red-shifted: UiO-66-NH$_2$–anisole absorbs up to ~520 nm, while –NNDMA and –DPA extend further, roughly to 700 nm. This trend aligns with the electron-donating strength of the substituents (–OCH$_3$ < –NHC$_6$H$_5$ < –N(CH$_3$)$_2$): stronger donors push the MOF's frontier orbitals closer together narrowing the gap.

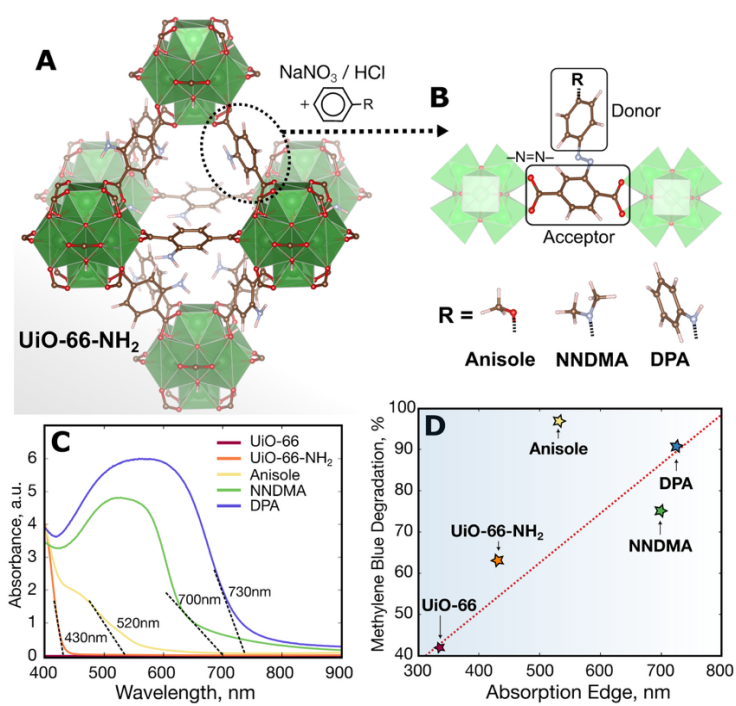

**Figure 1**. (A) Schematic structure of UiO-66-NH$_2$. Color key: carbon–brown; oxygen–red; hydrogen–white; nitrogen–grey; Zr$_6$O$_4$(OH)$_4$ metal nodes–green polyhedra. (B) General chemical structure of D-π-A-like linker functionalization, with an azo (–N=N–) bond as a π-bridge and different electron-donating groups R (R = anisole, NNDMA, and DPA). (C) Absorption spectra of dye-functionalized MOFs compared to unmodified UiO-66 and UiO-66-NH$_2$. The spectrum of UiO-66 is limited to the UV range (below 320 nm). (D) Methylene blue degradation versus absorption edge in tested MOFs.

Subsequently, we analyzed the photocatalytic performance of the dye-functionalized MOFs under broadband (UV–visible) illumination. Figure 1D illustrates the efficiency in the methylene blue (MB) degradation plotted against the absorption edge obtained from the Tauc plots. All three dye-functionalized MOFs show enhanced photocatalytic performance relative to unmodified UiO-66 and UiO-66-NH$_2$. Notably, however, the degree of photocatalytic performance enhancement does not strictly follow the extent of visible-light absorption. The UiO-66-NH$_2$–anisole sample, which had the most modest red-shift in absorption, achieves the highest MB degradation efficiency among the three. This lack of direct correlation suggests that more factors beyond the shifted absorption edge influence the overall photocatalytic efficiency.

**Insights on Electronic Structure and Charge-Separation from DFT and PL**

To analyze how each functional group modifies the electronic structure of the MOF and to check if this can help rationalize the experimentally observed trends, we carried out DFT calculations. As a benchmark, our DFT calculations for UiO-66-NH$_2$ yield a band gap of 2.86 eV (Figure 2A), in excellent agreement with the experimental value [9]. Next, we computed the electronic gaps of the isolated dye molecules (Figure 2A). The resulting HOMO–LUMO gaps of anisole, NNDMA, and DPA in the gas phase (calculated for the molecules in vacuum) are all significantly larger (by ~2 eV) than the band gaps of the corresponding functionalized MOFs deduced from the experimental absorption edges (cf. Figure 1C). This indicates that the observed visible absorption in the MOF samples is indeed due to the covalent integration of the dyes into the framework rather than residual free molecules.

Subsequently, we built structural models of dye-modified UiO-66-NH$_2$ and found that each functionalized linker can be accommodated in a pore of the MOF (cf. Supplementary Figure S3), although the substituents (especially the larger DPA and NNDMA moieties) do cause some local distortion of the surrounding framework. The phenyl rings of DPA and the dimethylamino group of NNDMA come in close contact with the pore walls causing slight shifts of Zr-node and linker positions (cf. Supplementary Figure S4), which could potentially lead to the loss of some neighboring linkers.

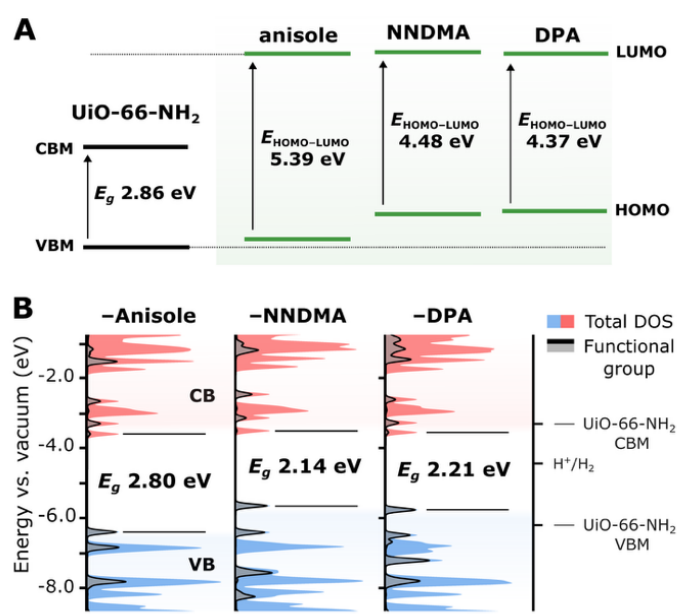

**Figure 2.** (A) DFT-calculated band gap of pristine UiO-66-NH$_2$ compared to the HOMO–LUMO gaps of the free dye molecules in vacuum. (B) Total and projected density of states (DOS) for the functionalized MOFs, showing the emergence of dye-localized states (dark grey area) at the valence band edge and their influence on band gap narrowing. Models assume one dye per pore and partially diazotized linkers bearing OH groups to reflect realistic synthetic conditions. All energies are aligned to the average electrostatic potential in the pore center (vacuum level).

Importantly, our structural models suggest that, due to steric hindrance, a maximum of one dye molecule can be anchored per pore, so the functionalization degree is inherently limited – only one of six linkers can be dye-modified at best. The remaining fraction of diazotized linkers is likely to undergo side reactions. A possible scenario is hydrolysis of the diazonium moiety (–N≡N$^+$) to form phenolic –OH on the linker.

This partial functionalization is in good agreement with photoluminescence (PL) spectra of UiO-66-NH$_2$ and modified structures illustrated in Figure S5 and S6. While the pristine UiO-66-NH$_2$ exhibit a single emission peak at 451 nm (with no additional features across the 375–845 nm measurement range), a dual emission character is observed in the dye-modified MOF samples. One PL peak is associated with a fraction of the parent framework, and the second one is attributed to the fraction subjected to the respective linker-functionalization. In UiO-66-NH$_2$-Anisole, the main peak is red-shifted by about 16 nm and broadened, and the second emission band appears near 600 nm, with an intensity approximately 20% lower than that of the main peak. The UiO-66-NH$_2$-NNDMA sample displays a 10 nm blue shift of the non-functionalized part along the with appearance of a broad, intense emission at 700 nm. In case of UiO-66-NH$_2$-DPA, the main peak remains largely unchanged save for slight broadening, while the additional feature appears near 715 nm.

This observation is indicative of partly loaded structures where large parts of the UiO-66-NH$_2$ structure remain non-modified. The shift and broadening of the UiO-66-NH2 PL peak can result from strain and structural defects introduced during the functionalization processes. Importantly, even partial functionalization is sufficient for enhanced light-harvesting and photocatalytic performance, as shown in Figure 1C,D.

To simulate partial functionalization in DFT calculations, the supercell models included five linkers per unit cell with –OH (from diazonium hydrolysis) and one linker is functionalized with a dye. The computed density of states (DOS) for the mixed-linker models illustrated in Figure 2B reveals that new occupied states appear above the valence band maximum (VBM) of the parent MOF. These states, primarily introduced by the HOMO of the anchored dye molecules, raise the valence band edge upward, hence reducing the gap (by ~0.7 eV for DPA and NNDMA and ~0.1 eV for anisole). The extent of the shift correlates with the donor strength: NNDMA and DPA (with the strongly donating –N(CH$_3$)$_2$ and diarylamine) introduce higher-lying occupied states than anisole (with the moderate –OCH$_3$ donor) [19]. This trend is fully consistent with the bathochromic shifts observed in the UV–Vis spectra (cf. Figure 1C).

Beyond band gap values, DFT provides crucial information on the spatial distribution of the frontier electronic states. Figure S7 illustrates the charge density associated with the highest occupied orbital (nominal VBM) and the lowest unoccupied orbitals (conduction band minimum, CBM) for the dye-modified MOFs. In all cases, the highest occupied state is mainly localized on the dye-functionalized linker – specifically, over the attached donor molecule and the azo bridge – while the CBM retains the character of unmodified UiO-66-NH$_2$ and is primarily distributed over the aryl and carboxylate moieties of the linkers, with partial contribution from the Zr oxide nodes.

This D–π–A type segregation of frontier electronic states is expected to facilitate intraframework charge transfer upon photoexcitation: When the dye-functionalized MOF absorbs a photon, the excited electron transfers from the electron-donating dye moiety to the framework's linker that acts as an acceptor, eventually reaching the catalytic sites (the Zr$_6$O$_4$(OH)$_4$ nodes of the MOF). The hole left at the dye segment can be then quenched without causing rapid recombination. To search for signatures of the photoinduced charge transfer we turned to continuous-wave electron paramagnetic resonance (cw EPR) spectroscopy.

**Probing Photoinduced Charge Transfer by EPR**

Before exploring the effect of photoexcitation, it was important to probe the presence of defects introduced during the functionalization, as these can potentially act as charge trapping sites. The EPR spectra of the MOF samples, collected at room temperature without exposure to light, are shown in Figure 3A. They reveal the presence of several paramagnetic species.

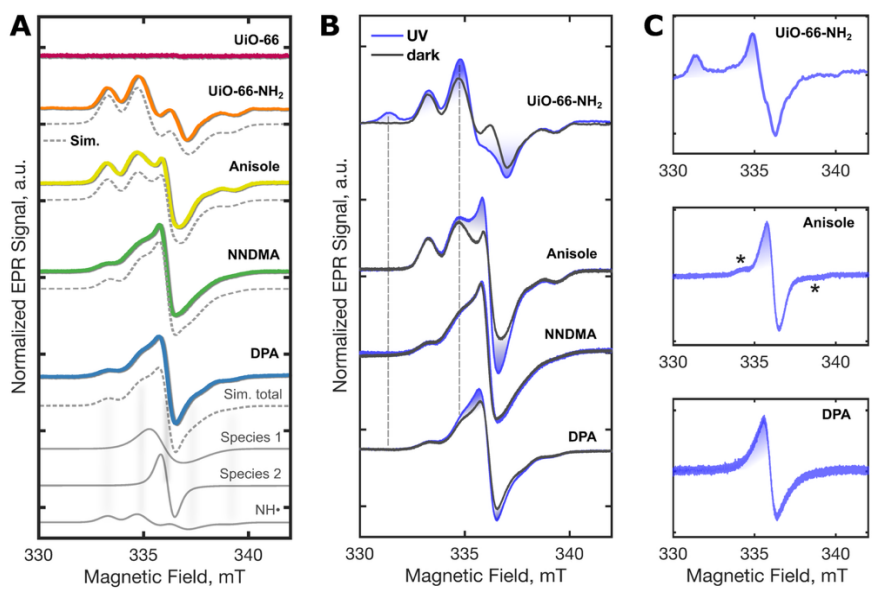

**Figure 3**. (A) EPR spectra (solid lines) measured at room temperature in the dark together with spectral simulations (doted). The parent MOF UiO-66 has no EPR signal, while we detect a strong •NH related signal for all modified MOFs. For DPA the individual signal contributions are shown in the lowest group of spectra. All spectra can be simulated as a superposition of the •NH group and two unspecified organic radicals, species 1 and 2. (B) Upon broadband UV-irradiation (blue traces) compared with the dark spectra (gray traces) additional signal intensity emerges. Field positions where the main photoinduced changes in the UiO-66-NH$_2$ EPR spectrum occur, are marked with vertical dashed lines. (C) Photoinduced signals obtained as a difference between the illuminated and dark EPR spectra.

By simulating the EPR spectra, we identified three primary radical centers. The lowest group of traces in Figure 3A shows an example of the deconvolution for the UiO-66-NH$_2$–DPA sample (see Supplementary Figure S2 for the simulations of the other functionalized MOFs). The first species is a nitrogen-centered radical, which is present in the parent UiO-66-NH$_2$ and attributed to an NH• group on the linkers [20]. It consists of four EPR lines due to hyperfine interaction between the unpaired electron and the nearest $^{14}$N and $^1$H nuclear spins. The observation of the NH• radical agrees with the observation of two PL peaks in Figure S5 and indicates the presence of non-modified UiO-66-NH$_2$ fraction. The other two contributions (marked as Species 1 and 2) that appear only for the dye-modified samples are both described with $g$-factors close to that of the free electron ($g_e$=2.0023). One contribution is a broad line best fitted assuming a slightly anisotropic $g$-factor ($g_X$=2.003, $g_Y$=2.003, $g_Z$=2.006), and the other one is a narrow EPR line at $g$=2.004.

The $g$-factor (or, more specifically, its difference from $g_e$) is a definitive spectroscopic fingerprint determined by the radical's electronic structure and chemical environment via spin-orbit coupling effects. Based on the observed $g$ values of Species 1 and 2, we can conclude that these are related to neither oxygen-centered radicals (such as O$^-$ and O$_2^-$ radical ions) associated with the $Zr_6O_4(OH)_4$ nodes nor to $Zr^{3+}$ defect centers, which typically exhibit an order of magnitude larger deviation from $g_e$ [7, 21]. On the other hand, $g$ values close to that of $g_e$ are typical for carbon (organic) radicals, which are likely to be generated during diazonium reactions. The absence of resolved hyperfine interactions signifies that the electron density of the radicals is significantly delocalized.

As a hint to understand the radical formation in these samples, we note that diazonium coupling is known to involve a free radical chain reaction yielding both radical intermediates (such as aryl) and stable radicals [22, 23]. Alternatively, nitrite ions accumulated in the pores near the MOF surface can react preferentially with the amines (NNDMA and DPA) to form N-nitrosamines [24, 25]. Subsequent homolytic cleavage of nitrosimes' N–N bonds can result in the formation of amino radicals [26, 27].

Importantly, the weights of the two new radical species (1 and 2) are relatively low in all three modified MOFs and comparable to that of the •NH radical. This indicates that the diazo-coupling reaction did not generate a high concentration of defects, and the framework's integrity remains intact. Among the three, the UiO-66-$NH_2$–anisole sample exhibits the smallest concentration of Species 1 and 2. This may suggest that the smaller anisole group bonds more cleanly with less perturbation to the structure, whereas the bulkier DPA and NNDMA introduce slightly more disorder (e.g. occasional incomplete grafting, side-reactions leading to radicals, and the generation of missing-linker defects [28–33]). Overall, the EPR characterization confirms that while a few radical species are present after functionalization, their concentrations are reasonably low.

Now to elucidate the photoinduced charge transfer mechanisms in the dye-modified MOFs, we investigated the formation of electron spin centers under *in situ* illumination. Photoinduced spin centers are a telltale sign of electron-hole separation associated with the MOF's photocatalytic activity [7]. Figure 3B shows EPR spectra in the dark together with those taken under constant UV-rich illumination (280–450 nm). Changes in the EPR spectra can be observed at several spectral positions. To better highlight these changes, Figure 1C shows the spectra with the dark contributions subtracted.

In the parent UiO-66-$NH_2$ MOF, the photoinduced EPR signal (Figure 1C, top panel) is well characterized and commonly attributed to a superoxide ($O_2^{-•}$) radical formed due to electron transfer to an $O_2$ molecule adsorbed at an uncoordinated Zr site [7, 8]. It exhibits a nearly axially symmetric g-tensor with the principal values $g_X = 2.002$, $g_Y = 2.009$, and $g_Z = 2.033$. Similar to the parent framework, the anisole-functionalized sample shows appreciable photo-EPR response, suggesting efficient electron-hole separation. At the same time, its photoinduced EPR spectrum is substantially different. It consists of a dominant line at $g \approx 2.004$, accompanied by broad, asymmetric satellite features near $g \approx 2.014$ and $g \approx 1.990$ (marked with asterisks in Figure 3C). Most likely, chemical conditions involved in diazo functionalization suppress the formation of node-bound superoxide in favor of another linker-centered or Zr-node-based radical.

As for the other two dye-modified MOFs, the DPA sample exhibited a considerably weaker photoinduced EPR signal, while NNDMA showed no discernible photo-EPR response (Figure 1C, middle and bottom panels). This was an unexpected result, since our electronic structure calculations suggest that all three dye-modified MOFs should exhibit efficient charge transfer from the electron-donating functional group to the framework. To seek for a possible explanation, we carried out time-dependent DFT (TDDFT) calculations accompanied by transient photoluminescence (trPL).

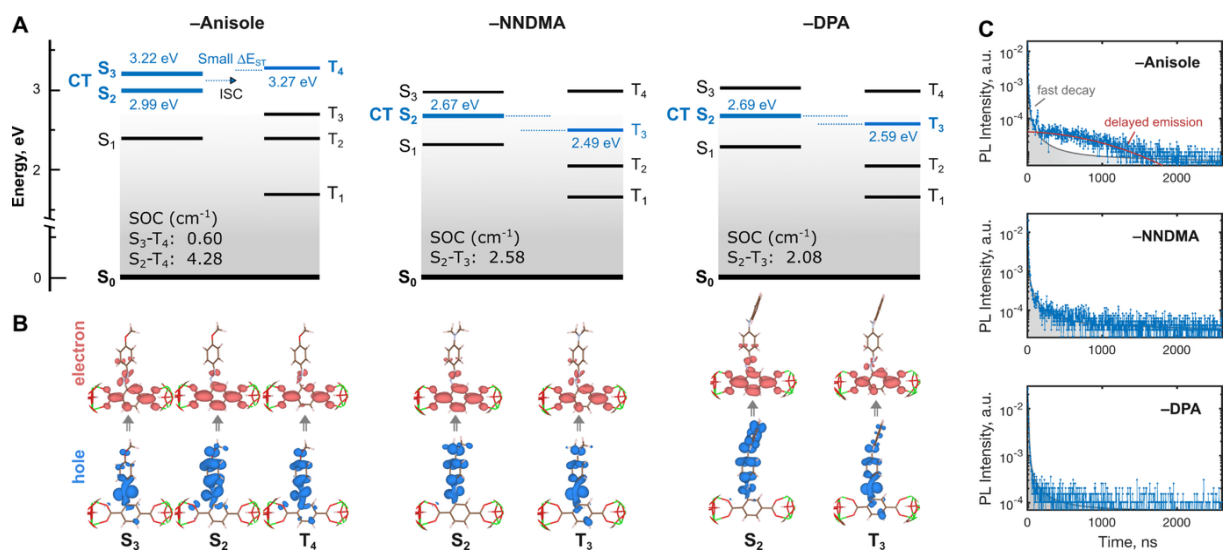

**Figure 4.** (A) TDDFT-derived excited-state structure of UiO-66–NH$_2$ modified with anisole, NNDMA, and DPA donors. The charge-transfer (CT) singlet and triplet states are highlighted in blue, and the calculated spin-orbit coupling (SOC) matrix elements between them are shown in each panel. (B) Electron (*red*) and hole (*blue*) components of the difference orbitals associated with the identified CT states. (C) Time-resolved PL traces showing a fast decay of singlet excitation. For the anisole-modified MOF this is followed by delayed emission from triplet states.

At the first glance, the TDDFT results look similar across all three dye-modified systems (Figure 4A and B). The lowest singlet excited states (S$_1$) are of localized exciton (LE) character associated with the –N=N– bridge, while the actual charge-transfer (CT) states that generate charge-separated spins sit between 2.6 and 3.2 eV (≈ 480-390 nm), which is still accessible by the excitation wavelengths used in our EPR experiments. At the same time, the anisole-functionalized system exhibits a CT triplet state T$_4$ that is nearly resonant with one of the CT singlets (S$_3$). Along with the sizeable spin-orbit coupling (SOC) connecting these states, this suggests an efficient intersystem crossing (ISC) pathway. Once the system reaches a triplet CT state, the charge-separated donor-acceptor pair (D•$^+$/A•$^-$) becomes spin-protected: recombination is spin-forbidden, allowing the carriers to persist longer, find trapping sites, and yield a detectable photo-EPR Asignal.

In the NNDMA and DPA systems, the calculations suggest somewhat larger singlet-triplet gaps ($\Delta E_{ST}$), which might be enough to suppress ISC and facilitate faster recombination of the charge carriers. The lifetimes of D•$^+$/A•$^-$ pairs in these MOFs can still be sufficient for enhanced photocatalytic activity compared to the parent MOF (cf. Figure 1D) but too short to yield a steady EPR signal at room temperature.

These conclusions agree with the trPL measurements shown in Figure 4C (for detailed trPL results, see also Figure S8). The main UIO-66-NH2 emission band exhibits a similar decay behaviour for all samples, confirming comparable decay dynamics of the host structure. However, the additional linker-related emission bands show distinct characteristics. The anisole-modified MOF sample displays a fast decay component ($\tau_1$ ≈ 0.4 ns) followed by a long-lived delayed emission extending up to 1500 ns, indicating phosphorescence from triplet excited states. This dual-component decay – fluorescence and delayed phosphorescence – is also observed for the parent UiO-66-NH$_2$ MOF (cf. Figure S7). In contrast, long-lived triplet emission is not observed for UiO-66-NH$_2$-NNDMA and UIO-66-NH$_2$-DPA samples, whose linker-related emissions are dominated by fast singlet-based radiative decay pathways.

Together the mechanistic insights from EPR, TDDFT, and trPL may explain why anisole functionalization leads to superior photocatalytic performance compared to more electron-donating DPA and NNDMA.

# Conclusion

Our results have shown that grafting D–π–A dye units onto the linkers of UiO-66-NH$_2$ via diazo coupling offers a viable route to extend light absorption and enhance photocatalytic activity. Functionalization with anisole, DPA, and NNDMA systematically modifies the MOF's electronic structure by introducing new occupied states near the valence band edge. DFT calculations confirm that these states enable spatial charge separation via electron transfer from the dye to the framework backbone, while preserving the integrity of the host. EPR spectroscopy confirmed that only modest content of defect species is present in the dye-modified samples (indicating the framework remains largely intact).

By combining EPR spectroscopy with trPL measurements and TDDFT calculations, we find that the anisole functionalization strikes an optimal balance. It facilitates efficient charge transfer along with ISC into the triplet manifold that protects the photoinduced charge carriers from fast recombination. Moreover, it introduces minimal structural defects (potential trap states), thus preserving high charge-transfer efficiency.

From a practical standpoint, the insights gained here provide guidelines for future design of D–π–A-functionalized MOF photocatalysts. We highlight that involving characterization techniques like EPR is imperative for diagnosing trap states, charge-transfer efficiency, and the nature of reactive intermediates. More generally, our work contributes to the growing toolbox for engineering MOF-based materials for solar energy conversion and environmental remediation, connecting concepts from molecular dye chemistry with the robust scaffolds of reticular frameworks.

# Methods

**Synthesis of Dye-Functionalized MOFs:** UiO-66-NH$_2$ was synthesized according to Ref. [34] and subsequently post-synthetically modified via a diazonium coupling reaction to attach the aryl donor groups, following a similar approach to Otal *et al.* [12].

**Diffuse reflectance spectroscopy:** Reflectance spectra were measured using a UV-visible scanning spectrophotometer (SHIMADZU UV-2101PC) attached to an integrating sphere. Sample powders were mounted into the integrating sphere using a quartz slide. Reflectance was transformed to absorbance using the Kubelka—Munk model.

**Photocatalytic Characterization:** Photocatalytic activity was evaluated using the degradation of methylene blue (MB) as a model reaction. A stock solution of MB (2 mM) was prepared by dissolving 0.0639 g of dye in 100 mL of distilled water. This solution was diluted 1:10 to yield a working concentration of 0.2 mM. For each test, 15 mg of MOF sample was dispersed in 10 mL of the MB solution in a Pyrex test tube equipped with a magnetic stir bar. Titanium dioxide ($TiO_2$, anatase phase, Alfa Aesar, surface area 45 m²/g) was used as a reference photocatalyst. Prior to illumination, all suspensions were stirred in the dark for 1 hour to assess dye adsorption. After this equilibration step, 3 mL aliquots were withdrawn, filtered through 20 μm nylon membranes, and the absorbance was measured at 664 nm to determine the initial MB concentration. The remaining suspensions were then irradiated for 1.5 hours under continuous stirring using a 300 W xenon arc lamp as the light source. After irradiation, aliquots were again collected, filtered, and analyzed spectrophotometrically at 664 nm.

**EPR Spectroscopy:** CW EPR was done on a Magnettech MS5000 outfitted with an Oxford ESR 900 He flow cryostat. The deviation of the set temperature was less than 0.2 K. Two types of light sources were used, a broadband Dymax BlueWave 50 UV lamp (280 nm to 450 nm). The power of the UV was measured considering the distance between the sources and the irradiated sample. Thus, the broadband UV lamp has a power of 5.1 mW, respectively. A microwave power of 1 mW was chosen for optimal signal-to-noise ratio of the main EPR spectra without saturation effects. However, experiments on saturation behavior have also been conducted (Figure S9). EPR spectral simulations were performed with the EasySpin software package [35] for MATLAB.

**Computational Details:** The atomic structures and electronic properties of both unmodified and dye-functionalized UiO-66-$NH_2$ were investigated using density functional theory (DFT) under periodic boundary conditions. All DFT calculations were performed with the Quantum ESPRESSO package v7.2 [36, 37]. The generalized gradient approximation of the PBE exchange-correlation functional [38] was used for geometry optimizations, with a plane-wave kinetic energy cutoff of 700 eV, Brillouin zone sampling at the Γ-point, along with PAW pseudopotentials taken from PSLibrary v1.0 [39]. Starting from the experimentally reported UiO-66-$NH_2$ crystal structure, we constructed supercell models incorporating one functionalized linker (with anisole, NNDMA, or DPA) per $Zr_6$ node (i.e., one dye per pore) along with five remaining linkers in a partially diazotized state (modeled as –OH substituents on what were formerly –$NH_2$ groups) to simulate incomplete reactions and potential side-products. The atomic positions were fully relaxed until forces converged below 0.01 eV/Å, while the lattice constants were kept fixed. Electronic structure calculations were then carried out using the screened hybrid HSE06 functional on the optimized structures to obtain accurate band gap estimates [40, 41]. To align the energy levels and compare band edge positions, we evaluated the electrostatic potential in the center of the MOF pores and took its average as the vacuum level reference, following the procedure of Butler *et al.* [42]. This was facilitated by the MacroDensity tool [43]. TDDFT calculations were performed with the ORCA program package [44] using cluster models cut from the DFT optimized supercells. The clusters consisted of one linker and two Zr-oxo nodes; the remaining (unsaturated) coordination sites of each Zr-oxo node were terminated by $HCOO^-$ anions. The TDDFT calculations employed the TPSSh exchange-correlation functional [45] and the def2-TZVP basis set [46] combined with the RIJCOSX approximation and the corresponding auxiliary basis set [47].

**Photoluminescence and time-resolved photoluminescence measurements:** For photoluminescence (PL) and time-resolved photoluminescence (trPL) measurements, powder samples were placed inside the 0.5mm thick metal frame sealed between two glass slides (transparent in the 320-2500 nm range) from both sides (Figure S1). The PL and trPL measurements were performed using an Edinburgh Instruments FLS980 spectrometer. The samples were optically excited with a pulsed 375 nm laser source.

# Author contributions

A.K. performed the EPR and photo-EPR measurements and simulations. V.V. carried out the photoluminescence and time-resolved photoluminescence measurements. A.K., V.V., and A.S. analyzed and interpreted the spectroscopic data. T.B. and W.G.S. performed the first-principles calculations and theoretical analysis. E.O. and K.T. conducted the chemical synthesis, photocatalytic experiments, and UV–Vis characterization. A.K. and T.B. conceptualized and coordinated the project. All authors contributed to the discussion of the results and to writing the manuscript.

# Conflicts of interest

There are no conflicts to declare.

# Data availability

All data supporting the findings of this study are available within the article and its Supplementary Information. Additional data related to this paper may be requested from the corresponding authors.

# Acknowledgements

A.K., A.S. acknowledge financial support from the Würzburg-Dresden Cluster of Excellence on Complexity and Topology in Quantum Matter ct.qmat (EXC 2147, DFG project ID 390858490). T.B. and W.G.S acknowledge the Paderborn Center for Parallel Computing (PC2) for the provided computational resources.

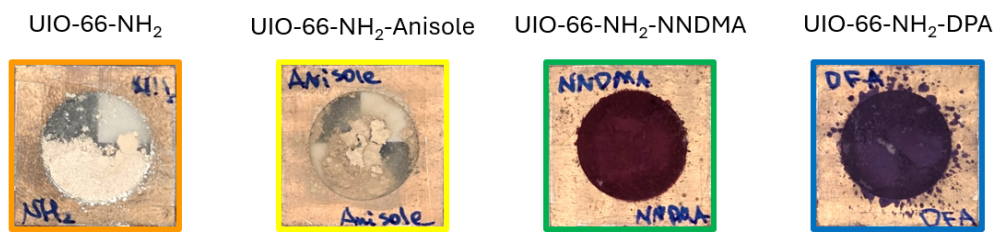

**Figure S1.** Optical photographs of pristine UiO-66-NH$_2$ and linker-modified (-Anisole, -NNDMA, and -DPA) powder samples mounted in sample holders for PL and trPL measurements.

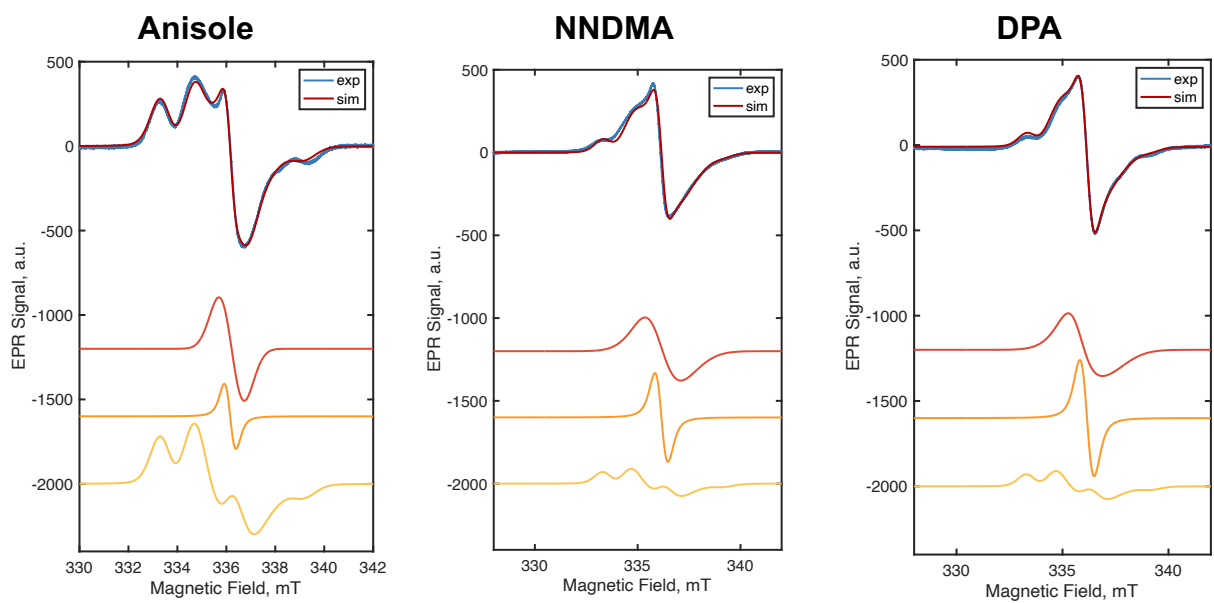

**Figure S2.** EPR simulations for dye-modified MOFs. The EasySpin parameters of the three superimposed spectral components are listed in Table S1.

**Table S1**. EasySpin parameters of the spectral components used in the EPR simulations in Figure S2

|   | Parameter | Anisole | NNDMA | DPA |
|---|---|---|---|---|
| Species 1 | g | 2.0042 | 2.0042 | 2.0042 |
|   | lw | [0.4, 0.4] | [0.5, 0.6] | [0.55, 0.6] |
|   | weight | 0.037 | 0.40 | 0.55 |
| Species 2 | g | [2.0025, 2.0025, 2.0066] | [2.0025, 2.0025, 2.0051] | [2.0025, 2.0025, 2.0066] |
|   | gStrain | [0.003, 0.003, 0.002] | [0.015, 0.015, 0.006] | [0.015, 0.015, 0.001] |
|   | lw | 1.0 | 1.0 | 1.0 |
|   | weight | 0.15 | 1.30 | 1.20 |
| NH• | weight | 1.00 for the EasySpin parameters of the NH• radical, see Ref. [20] of the main text | | |

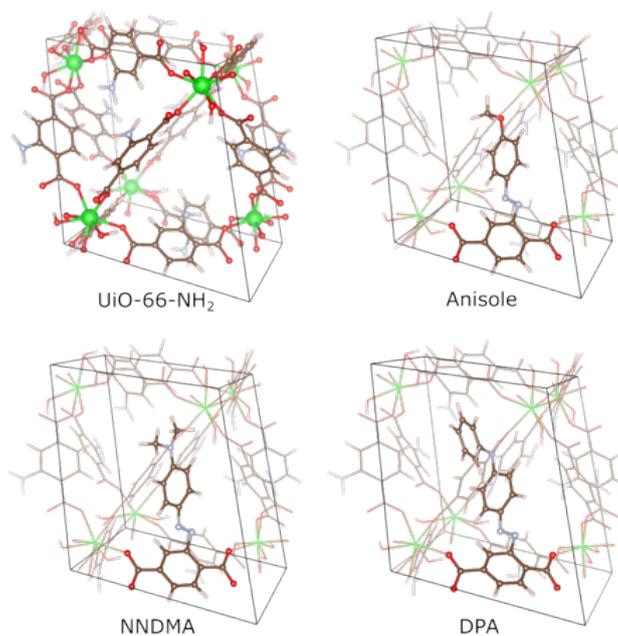

**Figure S3.** Schematic representation of DFT-optimized supercell structures of non-modified and diazo-coupled UiO-66–NH$_2$ MOF (color key: carbon, brown; oxygen, red; hydrogen, white; nitrogen, grey; and zirconium, green).

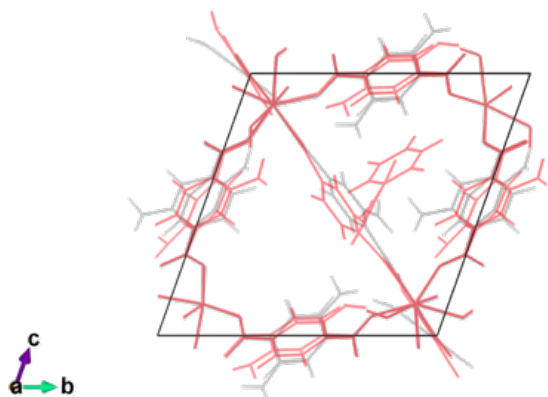

**Figure S4.** Overlay of the nonmodified (grey) and DPA-coupled (red) UiO-66-NH$_2$ DFT-optimized structures illustrating the distortions caused by the introduction of the functional group.

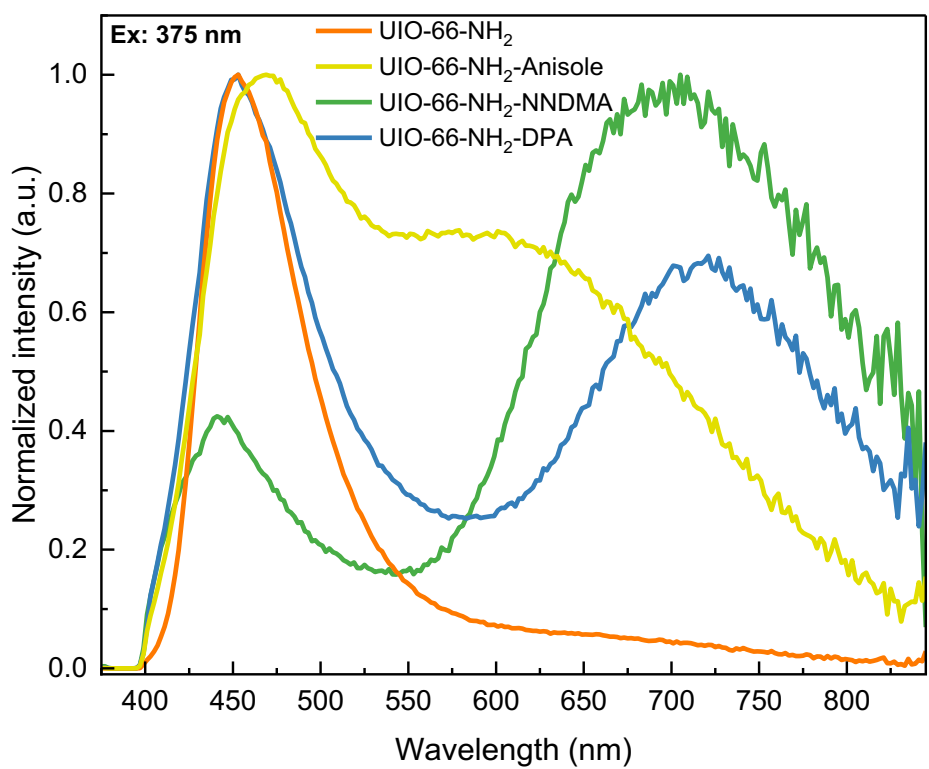

**Figure S5** Normalized PL spectra of pristine and modified UIO-66-NH$_2$ in the 375 – 845 nm range. The dual emission character of the loaded structures is indicative of incomplete loading by diazo-functionalization. Significant part of the UiO-66-NH$_2$ remains non modified.

The UIO-66-NH$_2$-Anisole sample exhibits a ~16 nm red shift of the main UIO-66-NH$_2$ peak from 467 nm, accompanied by peak broadening. An additional broad emission band appears near 600 nm, with an intensity approximately 20% lower than that of the main peak. UIO-66-NH$_2$-NNDMA displays a ~10 nm blue shift of the UIO-66-NH$_2$ characteristic emission peak to 442 nm, along with the appearance of a broad, intense emission at 700 nm, roughly twice the intensity of the 441 nm peak. The UIO-66-NH$_2$-DPA shows retention of the main emission peak at 451 nm, but with slight peak broadening and an additional broad feature near 715 nm, ~25% lower in intensity than the 451 nm emission. PL results demonstrate that D–π–A linker incorporation strongly affects the optical transitions in UiO-66-NH$_2$, leading to new emission channels and spectral shifts. A detailed graph of peak shifts is shown in Figure S6.

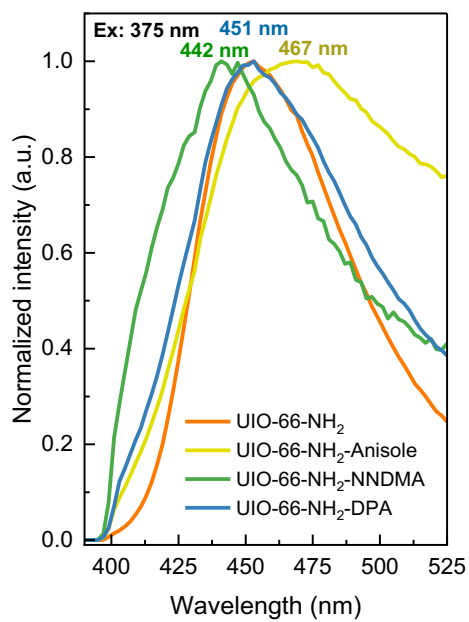

**Figure S6.** Normalized PL spectra of pristine and modified UIO-66-NH$_2$ in the 380 – 525 nm range.

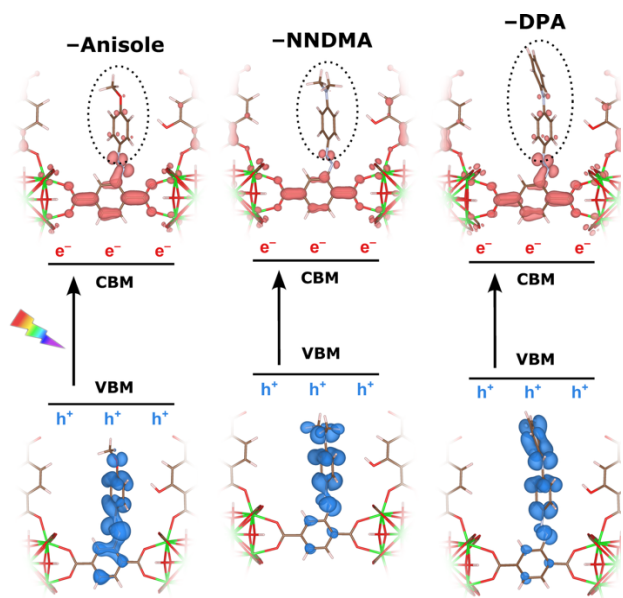

**Figure S7**. Visualization of the highest occupied and lowest unoccupied states in dye-modified systems, showing spatial separation of the photoexcited electron (mainly on the linker backbone) and the hole (localized on the dye and azo bridge).

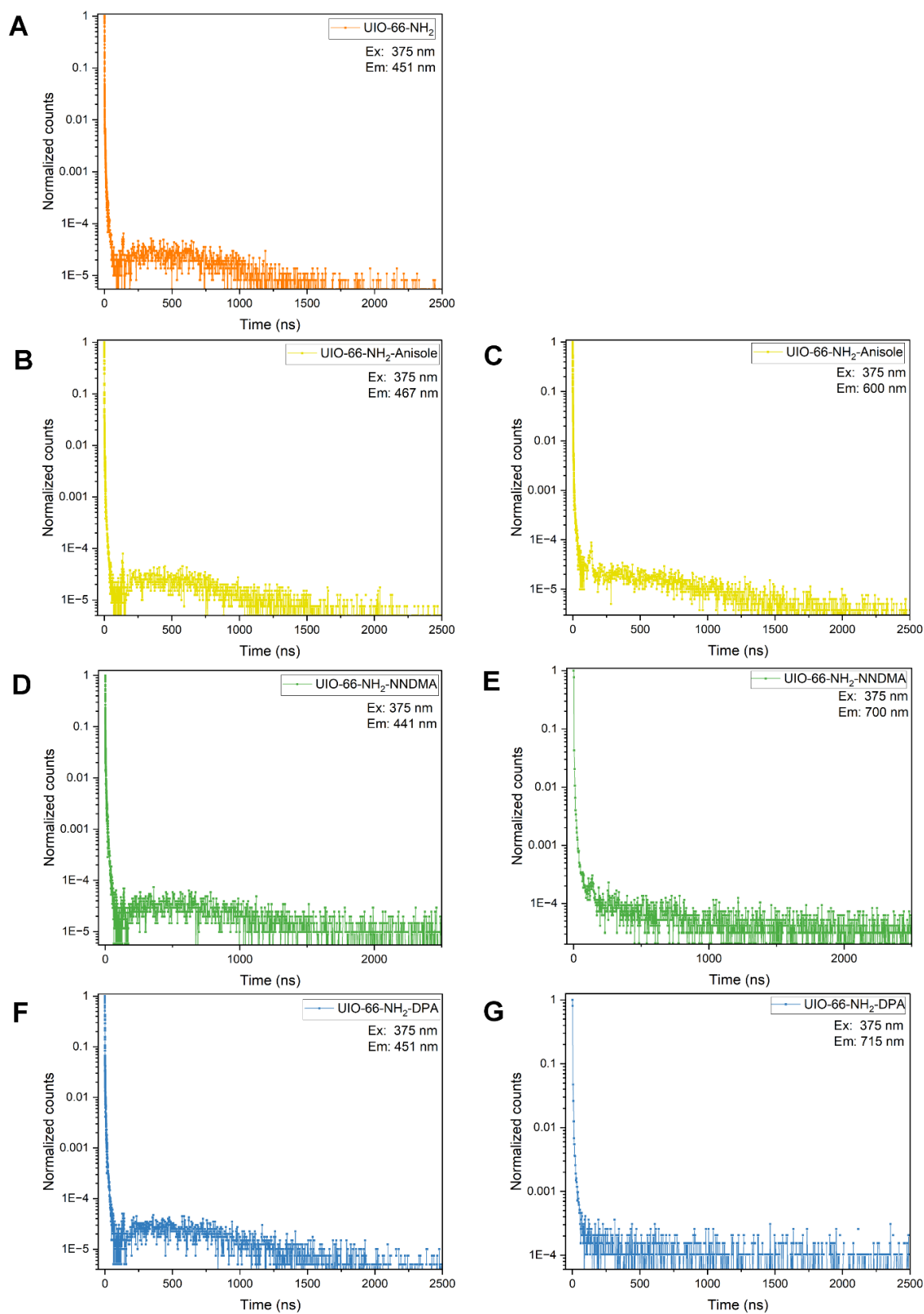

**Figure S8.** Time-resolved PL decay of (A) UIO-66-NH$_2$ at 451 nm; UIO-66-NH$_2$-Anisole at (B) 467 nm and (C) 600 nm; UIO-66-NH$_2$-NNDMA at (D) 441 nm and (E) 700 nm; UIO-66-NH$_2$-DPA at (F) 451 nm and (G) 715 nm emission wavelength. Since the dye-modified samples exhibit two PL peaks (cf. Figure S5) – 1$^{st}$ originating from the fraction of UiO-66-NH$_2$ parent framework and 2$^{nd}$ related to fraction with the respective linker modification – both regions were probed separately to analyze their individual recombination behaviors.

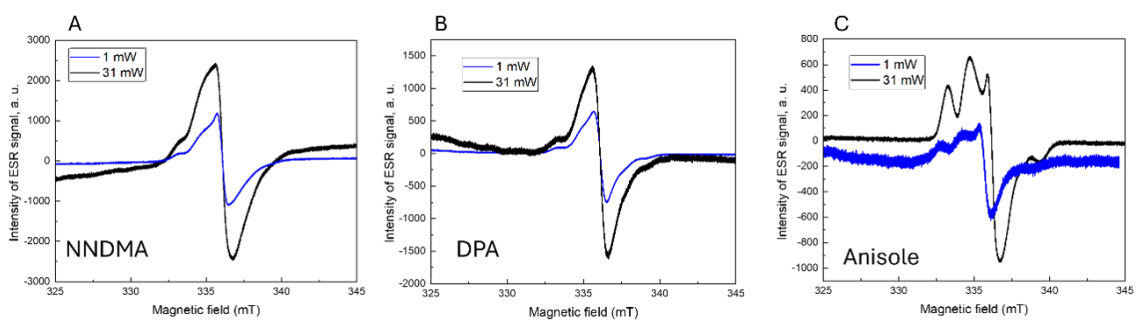

**Figure S9.** The microwave power dependence of EPR signals for dye-functionalized MOFs at the 293 K. The almost identical behaviour under two different microwave power settings displays the same spin-spin and spin-lattice relaxation times for the different species contributing to the superimposed spectra.

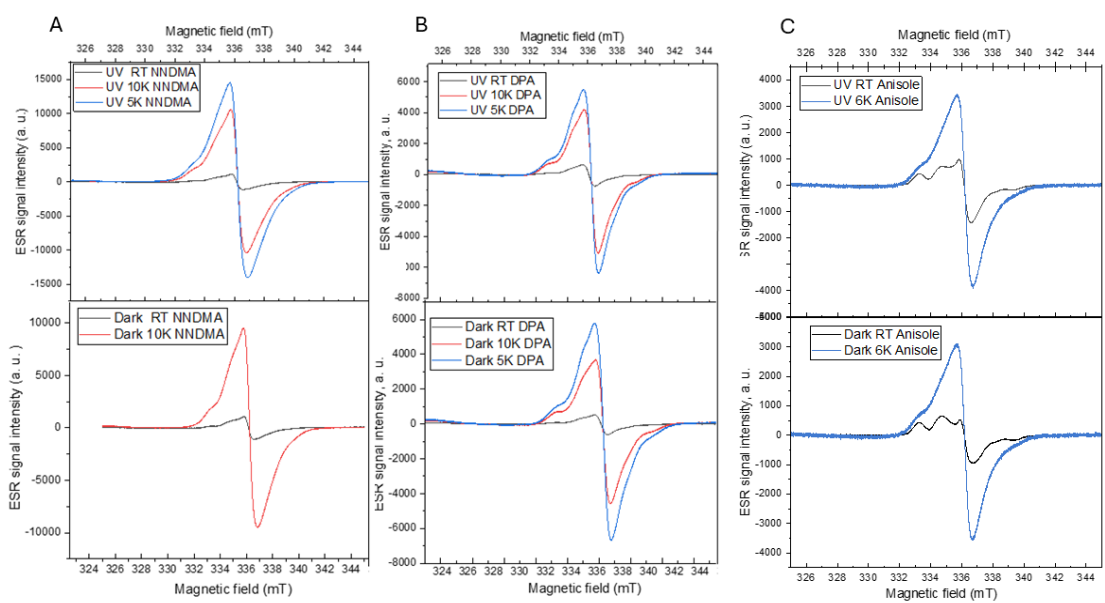

**Figure S10.** The temperature dependence of EPR signals for dye-functionalized MOFs at 1 mW microwave power. The temperature dependence demonstrates a homogeneous increase in the intensities of the spectra for NNDMA and DPA modifications indicating relatively long relaxation times of the paramagnetic centers which are not significantly temperature-dependent. In the case of the Anisole-modified spectra at room temperature, the spectrum from ·NH groups dominates, but the low-temperature spectrum is identical to the modifications described above. This confirms the formation of radicals of the same nature in all the modified samples.